**Founding the First Chemistry Laboratory in Russia: Mikhail Lomonosov's Project**

Robert P. Crease and Vladimir Shiltsev[1]

*Abstract:* This article, the third in a series about the Russian scientist Mikhail Lomonosov (1711–1765), covers the first decade of his research at the St. Petersburg Academy of Sciences, from his return from an educational his trip abroad in 1741, to the mid-1750s. Lomonosov's major focus was on the establishment of the first Russian laboratory used to introduce modern experimental chemistry and physics methods both to original research and education. The lab supported studies of the physics of colors, chemistry and physics of glasses and training of the Academy students. This article describes how Lomonosov, first an Adjunct Professor and then as a young Professor, fought to create the chemistry lab, and then to establish a broad program of experiments and tests there. The construction of laboratories to be used not just for research but also education only became widespread in the early 19th century, but Lomonosov's laboratory had a significant impact on the early development of the Academy and on Russian science.

---

[1] Robert P. Crease (corresponding author) is Professor of Philosophy at Stony Brook University. Vladimir Shiltsev is Professor at Northern Illinois University.

**Introduction**

In August 1748, the St. Petersburg Academy of Science began construction of a chemical laboratory on Vasilievsky Island in St. Petersburg under the supervision of Mikhail Lomonosov, a professor at the Academy. It is often cited as a key moment in the history of Russian science. But why exactly? Chemical research was carried out previously in Russia. Among other developments, Peter the Great had established or supported numerous laboratories of various kinds about four decades before. Furthermore, the Academy's chemical laboratory made no earth-shattering discoveries in chemical history, nor developed any revolutionary techniques. It was torn down after a few decades. This article is about why and how the laboratory was built, what made it unique, and what it can tell us about the symbolic role of laboratories in the history of science.

**Lomonosov's Chemical Education**

Lomonosov is one of Russia's most mythologized figures [Fig. 1]. Regarded as Russia's first scientist in the modern sense of the term, he was born in 1711 on Kurostrov Island, near the town of Kholmogory in the far north of Russia along the White Sea.[1] He traveled 600 miles south to Moscow in 1730, getting himself admitted to the prestigious Slavic-Greek-Latin Academy. Lomonosov then had three fantastically lucky breaks in a row. In the first, he was sent for further studies to the St. Petersburg Academy of Sciences ("the Academy"), established by Peter the Great in 1725. Lomonosov therefore got jump-started into a science education along Western lines. He was only at the Academy a few months when he had his second fantastically lucky break: along with two other students he was ordered to Germany to continue his studies. They were sent to learn mining, in a way that would bolster the Siberian expeditions planned by Peter the Great to discover and evaluate the mineral resources thought to be found in Russia's eastern regions. Lomonosov's third lucky break was that his education in Germany did not work out as anticipated, and the three Russian students ended up with an education considerably broader than the Academy planned, in a way that would foster the eventual chemical laboratory.[2]

The Academy initially intended to send the three students to Freiberg, Saxony to study with Johann Henckel (1678-1744), a prominent authority on mining and mining practices. But the students ended up studying first in Marburg with the great philosopher Christian Wolff to get them up to speed on the German language and on basic natural philosophy. As it happened, Wolff taught the new rationalist materialist science of Descartes and Boyle mixed with a strong dose of Leibniz. Wolff in particular was a promoter of the corpuscular philosophy, a modern approach to understanding the nature of matter.

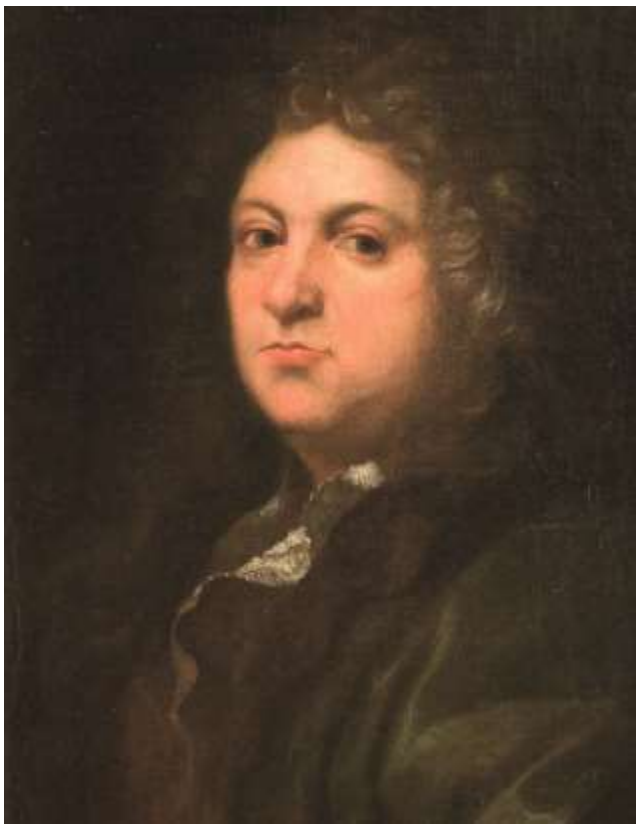

Fig.1: Portrait of Mikhail Lomonosov (1711–1765) in his 30s, around the time of construction of the first chemical lab. The portrait (author and attribution unconfirmed) is in the Russian National Library in St.Petersburg. Source: N. P. Kopaneva ''Mikhail Vasilievich Lomonosov: k priumnozheniyu pol'zi i slavi Otechestva'' [''Mikhail Vasilievich Lomonosov: to Motherlands' benefit and honor''], *Nauka iz Pervikh Ruk [Science First Hand]* **4** (40) (2011), 21–43, website http://www.sciencefirsthand.ru/pdf/sfh_40_20-43.pdf

Corpuscular philosophy was a way of breaking free from Aristotle without being atheistic. Aristotle had approached matter as arising from natural forms imposed on matter. The intuition of Boyle and other natural philosophers at the end of the alchemical era, who influenced Lomonosov via Wolff, was that a better way to understand the structure of matter was to approach them not top-down, so to speak, through forms, but bottom-up, as consisting of pieces of matter. The French called these pieces of matter *corps*, so this approach acquired the generic name 'corpuscular philosophy.' Corpuscular philosophy regarded the nature of matter as best investigated by determining the size, shape, and motion of these bodies, and how they combined with each other – and that these combinations give rise to all the properties of different kinds of matter. Different natural philosophers had different ideas about corpuscles, regarding them as material or immaterial, divisible or indivisible, simple or composite. But these philosophers shared the insight that the key properties of matter are a function of bodies, not of forms. This meshed nicely with alchemical practices, and provided a philosophical basis for experimentation, data collection, comparison, and analysis. It gave chemical practices a coherent systematic basis amenable to mathematical treatment.

Wolff provided Lomonosov and his fellow students with an appreciation for corpuscular philosophy, and an experimental take on math, physics and chemistry. Their principal chemistry textbook was Herman Boerhaave's *Elementa Chemiae*, published in 1732 and regarded as the first chemistry textbook; it had a practical focus, discussed assaying techniques, and had no discussion of phlogiston. Their principal model for an experimental chemist was Robert Boyle (1627-1691), regarded as the first modern chemist. Thanks to Wolff, the three Russians studying in Germany picked up the analytical and mathematical ideas, as well as practical methods, of Boyle and Boerhaave as interpreted and systematized by Wolff.

After studying with Wolff for three years, the Russian students finally went to Bergmaster Johann Henckel in Freiberg, pupil of famous chemist Georg Stahl,[3] with whom they learned the traditional assaying methods and mining techniques – exactly what they had been sent to acquire. The education that Lomonosov had received from Wolff provided him with a standpoint from which he could see the limitations of the assaying practices and other chemical techniques that he was learning, and a vision of what might take place in an advanced chemical laboratory. Henckel's laboratory was not a place for conducting research of the sort that Wolff had intimated was possible.

In mid-1741, Lomonosov left before the other two students. On his way home, among his other experiences he visited Johann Andreas Cramer, another *bergrat* (mining inspector) and metallurgy engineer, who was young, already famous, and had an excellent modern assaying lab. Lomonosov finally arrived back in St.Petersburg in the fall of 1741, and found the Academy's chemistry facilities woefully inadequate compared to what he had seen in Germany. The Academy had workrooms and laboratories, but its research had been backward compared to European laboratories from the beginning. The first Chair of chemistry was an Academician named Michael Burger, a German from Courland who had arrived St. Petersburg March 1726 and lasted a mere four months. After having too much to drink at a party at the home of the head of the Academy, Laurentius Blumentrost (1692-1755), Burger fell and died from injuries. Burger was succeeded by a naturalist, Johann G. Gmelin (1709-1755). Gmelin was a German botanist who had followed his Tübingen professors Bilfinger and Duvernoy to St. Petersburg, and had been at the Academy since the beginning. Gmelin became an Adjunct in 1727, and Academician (professor) of chemistry and natural history in 1731, but left St. Petersburg in 1732 for the Siberian expedition with Delisle and Müller – the expedition expected to join the one led by Vitus Jonassen (Ivan Ivanovich) Bering, which had already left. Gmelin remained at the Academy until 1743, though on his return he occupied himself with the Academy's plant collection. Gmelin left the Academy to return to Germany in 1747. With his departure, the Academy's already feeble chemical research ground to a halt.

Lomonosov and his two companions had been told that when they returned to the Academy they would each be given an extraordinary professorship in chemistry, which meant roughly what we might call an adjunctship. But Lomonosov, the first to return, was not granted it. The Academy's senior staff was suspicious of him due to the fact that he had left Germany early, and though highly regarded had quarreled with many of his German hosts. Nevertheless, Lomonosov began to write about chemistry, and completed a treatise called *Mathematical Principles of Chemistry*, which expressed the corpuscularian view that he had learned from Wolff. Lomonosov also began preparing a plan for building a modern chemistry laboratory at the Academy. It took advantage of Wolff's lectures about the possibilities of research, and Lomonosov's acquaintance with the assaying and mining laboratories of Henckel and others.

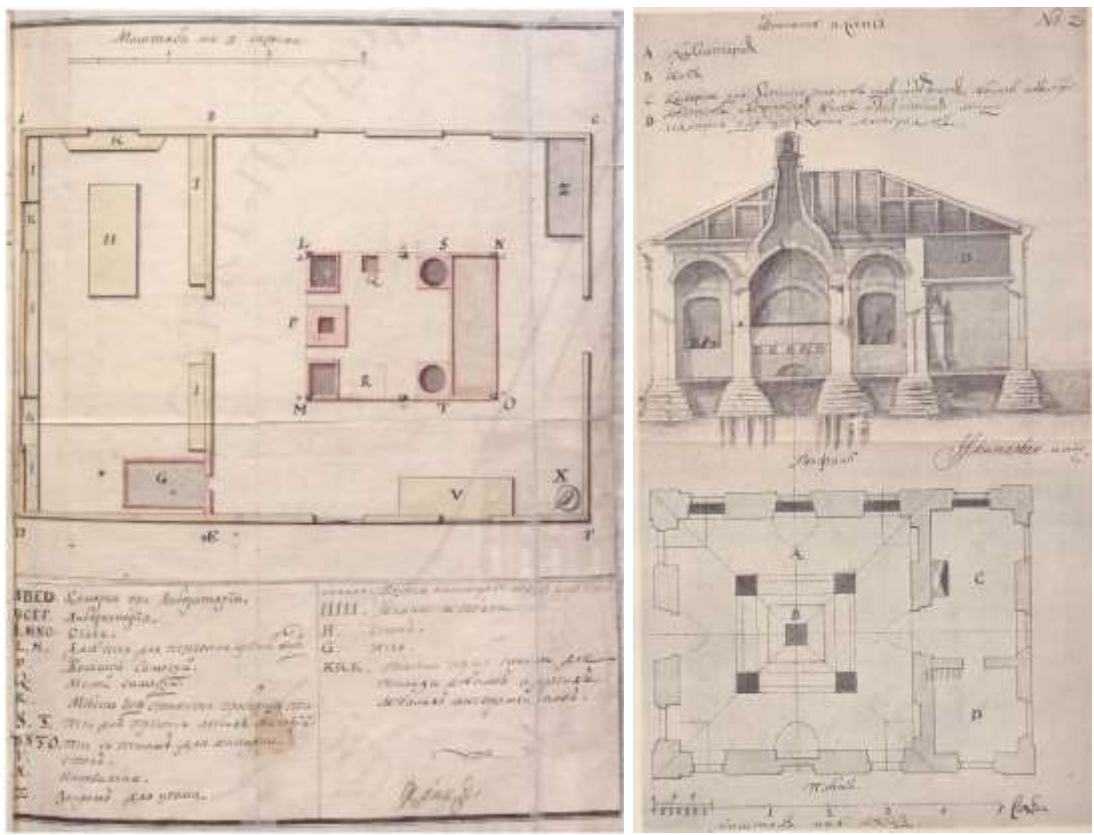

Fig.2: (a-left) Lomonosov's plan of the laboratory, May 1745. Scale at the top is in *sazhen's*, 1 *sazhen'* is 84 inches or 2.13m, so approximate outside dimensions of the lab are 12.4 x 8.4 m. *ABED* – auxiliary space, *BCEF* - laboratory spaces, *L,M* – furnace, *P* – two ovens for distillation of strong acids, *Q* – enforced-air oven, *R* – place for assay oven, *S,T* – ovens for distillation of light chemicals, *SNTO* – oven with sand for evaporation, *V* – table, *X* – anvil, *Z* – coal storage, *aaaa* – hood support ribs, *iiii* – shelves and boards, *G* – table, *H* – oven, *KKK* – under-window shelves. (b- right) Architect I.A.Schumacher's plan of the Lomonosov's chemical lab (1748) – front cross-section and room plan. *A* – lab space, *B* – oven, *C* – room for "…recoding of the test results, storage of small tools and books, and for delivering lectures", *D* – storage space for materials. The scale at the bottom is in *sazhen's*, so approximate outside dimensions of the lab are 14.1 x 10.8 m. (St.Petersburg Branch of the Archive of Russian Academy of Sciences.)

**Early Russian Laboratories**

Laboratories had been part of the Petrine vision for modernizing Russia from the beginning. Peter the Great took notes on the chemistry of the day, was familiar with many of its symbols and terminology, and knew some basic methods of ore assaying. He acquainted himself with chemistry textbooks, wanted to have some of them translated into Russian, and met with a chemist when he visited the Paris Academy of Sciences in 1717. Realizing the importance of a chemical industry to his vision of a modern Russia, he created laboratories for such things as testing ores and chemicals including gunpowder; testing and developing paints and dyes for boats, buildings, and textiles; producing glass, dyes, and pitch; and producing medicines.

"The first chemical laboratories were organized for the purpose of helping the metallurgical, chemical, military, and textile industries," writes the historian of chemistry P.M. Luk'yanov. "We do not know of any scientific investigations which were carried out in these laboratories."[4] Peter seems to have established the first sometime before 1707 at his residence in Preobrazhenskoie, in the vicinity of Moscow. This laboratory focused mainly on products for military use, such as gunpowder and rockets, as well as fireworks for Imperial displays. Other laboratories to produce ores, materials for military use, and medicines were soon built in Smolensk, St. Petersburg, and Moscow. The first laboratory for chemical research for which archival documents exist is the Berg-Collegium in St. Petersburg, created by Peter in 1720. According to Luk'yanov, Peter may well have carried out some simple analyses of ores himself at one of these laboratories. Laboratories with an applied industrial function were also known to exist in factories. But these were apothecary–like labs or industrial labs.
Private laboratories existed elsewhere, mainly to support the curiosity of the rich and enlightened;[5] for instance, Liebig, Lavoisier, and Boyle had their own laboratories, while Émilie du Châtelet and Voltaire built one at their Chateau de Cirey around 1734. But few private laboratories of that kind existed in Russia, and when they did it was to support curiosity of only few, personally rich and enlightened[6]. For example, Chancellor Count Alexey Bestuzhev-Ryumin (1693-1766) used his lab to develop "life drops" (*tinctura tonico-nervina Bestuscheffi*), an alcohol-ether solution of iron and a half-chloride, which was used to treat many diseases, from epilepsy to clogged vessels. Later on, Terentii Voloskov (1729-1806), successful amateur chemist and optician who made fortune in the dye industry, organized a lab in the city of Rzhev, while Vice-Chancellor of Berg-Collegium (Russian Ministry of Mining) Apollo Mussin-Pushkin (1760-1805) had another one in St. Petersburg.

As we show below, the chemical lab built by Lomonosov in 1748 was far different from all the above. First, his laboratory was supported directly by the state; its construction was supported by the Senate or Cabinet of Her Majesty Empress, while its operation was under the direction of the Academy, which in turn got funds from the State. Second, Lomonosov's lab was specifically designed for basic research and development of new knowledge, informed by the corpuscular approach that Lomonosov picked up in Marburg, which was amenable to systematic and mathematical treatment as well as to creating practical things like glass, mosaics, and assays using the mining and chemical instruments that he saw in Freiberg with Henckel. Third, it took the pioneering step of engaging students in empirical research on a large scale. This manner of teaching eventually became standard in modern science thanks to the early-19$^{th}$ century efforts of Friedrich Stromeyer (1776-1835) in Göttingen, Thomas Thomson (1773-1852) in Glasgow, and Julius von Liebig (1803-1873) in Giessen [7].

**Lomonosov's First Attempts**

While the Academy had some basic equipment for physico-chemical studies, opportunities for experimental research in chemistry were close to zero. For example, the only record of Lomonosov's early studies was that he and another adjunct, C.-E. Gellert (1713-1795), in May 1744 performed experiments in the Academy's physics cabinet (lab) on the action of nitric acid on metals in vacuum. They found that "strong vodka when dissolving metals without air acts differently than in the air." The next year, when asked by the Cabinet of Her Highness (the deputies of Empress Elizaveta Petrovna in charge of many state factories and manufacturing sites) to perform relatively simple and straightforward comparative chemical evaluations of seven different salts and mica, Lomonosov explained that he did not have the necessary tools and chemicals, and requested (and received) 30 rubles from the Academy for that purpose. Even the powerful Chancellor (formal head) of the Academy, Johann Daniel Schumacher, in 1745 admitted that "no chemical laboratory has been established to this day, and I must confess that at the Academy, no science had such poor success as this one"[8]. Lomonosov's first project at the Academy was to address that.

In January 1742, only a few months after his return, Lomonosov petitioned the Academy. Chancellery to build a chemistry laboratory (the text of this note has not been preserved). He submitted a second request in May 1743 to the Academic Chancellery. In it he wrote that "back in January of the past 1742, I, the lowest, filed a proposal to the Academy of Sciences to establish a Chemical laboratory, which did not exist yet even at the Academy of Sciences, so that I, the lowest, could work in chemical experiments for the benefit of the fatherland". In his report Lomonosov pointed out that the organization of such a laboratory is necessary so that he can "not only carry out chemical experiments for the growth of natural science in Russian empire and write about that in journals and dissertations in Russian and Latin, but at the same time … teach others physics, chemistry, natural mineral history … in order not to lose the amounts spent by H.I.H. [Her Imperial Highness] on my education in Germany spent and my labors invested."[9] Thus Lomonosov's idea for the lab was already different from that of the Petrine laboratories, for he conceived it as a place, not just for assaying and analysis, but also for research and teaching. Lomonosov identified two students – Stepan Krasheninnikov (1711-1755, future Academician and explorer of Kamchatka peninsula) and Alexei Protasov (1724-1796, future Academician in anatomy) – to whom he would like to teach "chemical theory and practice and also physics and natural mineral histories with all possible diligence." The petition was registered with the Academy Chancellery, which rejected it on the grounds that the Academy had no funds and the staffing possibilities for the lab were unclear[10].

Lomonosov made his third attempt to organize a chemical laboratory in March 1745. "The Academy of Sciences can clearly see," he wrote in this petition, "what a great and necessary means for studying nature and increasing art it is missing without the Chemical Laboratory. And although I have a zealous desire to practice in chemical works and bring honor and benefit to the fatherland, without a laboratory, I am forced to be content with only reading chemical books and theory, and will be forced to stop practical work almost completely."[11]. This time, M. V. Lomonosov attached a plan for the laboratory to his request [Fig. 2a] indicating its size and prospective work: to combine chemical and physical research; to synthesize pure chemicals and study the reactions between them; to analyze and synthesize chemical compounds; to obtain new substances needed in industry and nationally; and to enhance student education. Like the previous ones, this request was not approved. The minutes of the Academic Conference indicate, however, that Lomonosov's petition excited general discussion among academics about chronic shortage of funds and the need to submit a request to the (Governing) Senate [12].

That August, Lomonosov finally became a professor at the Academy – the first Russian-born academician, along with Trediakovsky – and in October submitted another request to the Chancellery, essentially repeating his previous arguments. This time the academicians unanimously supported their colleague, and on behalf of the Academic Conference, the project was sent to the Governing Senate:

> "Since we generally saw that the chemical laboratory at the Academy of Sciences is very much necessary for the study of natural things, without it, a professor of chemistry without it cannot thrive, any more than can a professor of astronomy without an observatory and related instruments. We most humbly ask the Governing Senate that, following the example of other
> 
> many glorious academies, to order construction of a chemical laboratory at the Academy of Sciences according to the attached drawing, and to furnish it with the necessary tools and other accessories, and for that to determine a special amount in excess of the one allocated to the Academy of Sciences. For this would be much more durable and safer if the aforementioned laboratory is ordered to be built of brick with vaults, and with a house for a professor of chemistry nearby, for it often happens that chemical operations continue for several days without interruption, and that professor needs to be there all the time. December 15, 1745 "[13].

Signees were Academicians (Professors) Delisle, Gmelin, Weitbrecht, Müller, LeRoy, Richmann, Trediakovsky and Lomonosov. The last request, for the lab to be close to the chemistry professor's residence, was quite important. the Academy was located on Vasilievsky Island, and many professors living in other parts of St. Petersburg had regular access problems, such as missing Academic conferences, at the times of ice formation (fall) and breakage (spring).

On July 1, 1746, Empress Elizaveta Petrovna signed a decree (ukase) to the Senate to construct a chemical laboratory at the Academy of Sciences "as in the attached drawing … at the expense of the Cabinet"[14].

**Construction**

The lab's location was determined by the need to have thechemistry professor's house nearby. Lomonosov had lived for many years (1741-1757) in the so-called "Bonov house" – a spacious single story building with 8 living rooms. Two other families also lived there at the time, that of the Academy's doctor and director of the botanic garden Johann Georg Siegesbeck (1686—1755,) and of the gardener Schturm, as well as rooms for servants.. The Bonov house was a former property of Peter I's Governor-General Herman Jensen Bohn (1672-1743), and was located next to Academy's Botanical Garden (current address, Vasilievsky Island 2$^{nd}$ Liniya  41-45 and 1$^{st}$ Liniya 52-56). The house and the plot had been the property of the Academy since 1739. At the end of June 1747, the Chancellery of Constructions asked the Academy to select the location of the chemical lab, and Lomonosov, along with a representative of the Constructions Bock as well as Ensign Shestakov, chose a plot in the courtyard of the Bonov House, thus next door to Lomonosov's residence (about 30 m from his door), bordering the botanical garden.

In March 1748, the Chancellery of Constructions notified the Academy that no funds were available to begin construction. But the Academic Chancellery asked an architect (Johann Jacob Schumacher, the brother of Johann Daniel, the boss of the Academy) to work with Lomonosov to update the cost estimate (1470 rubles 95 copecks) and design of the lab (see Fig. 2b).  On May 31, the "Sankt-Petersburg Vedomosti" newspaper published an announcement opening the bidding for construction of the lab. The announcement was published in several successive issues, and the public bidding took place in the Academy on July 28. The winning bid, for 1,344 rubles, was awarded to contractor Mikhail Gorbunov.   The lab was built of stone, about 14 meters long, 10.8 meters wide, and some 5 meters high. It had a hearth (a wide chimney) in the middle to serve several chemical furnaces, a flue, underground air ducts, one room for storage and another an auditorium for lectures, chemical balances, analyzing and writing up results [Fig. 3a].  In the meantime, Lomonosov began collecting and purchasing materials and equipment for the lab.

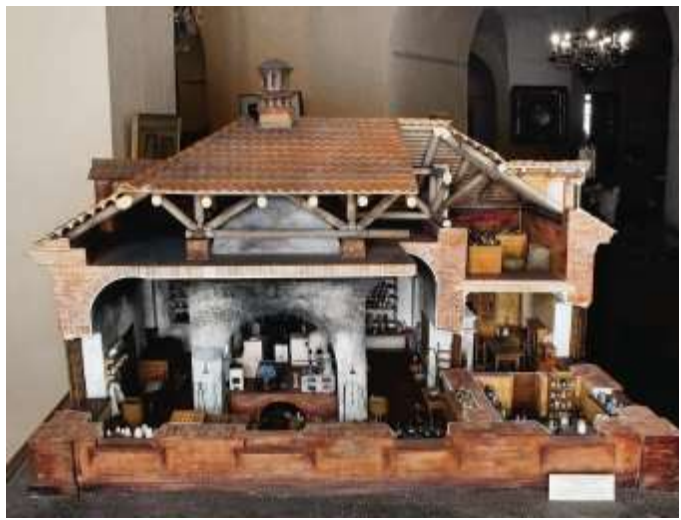

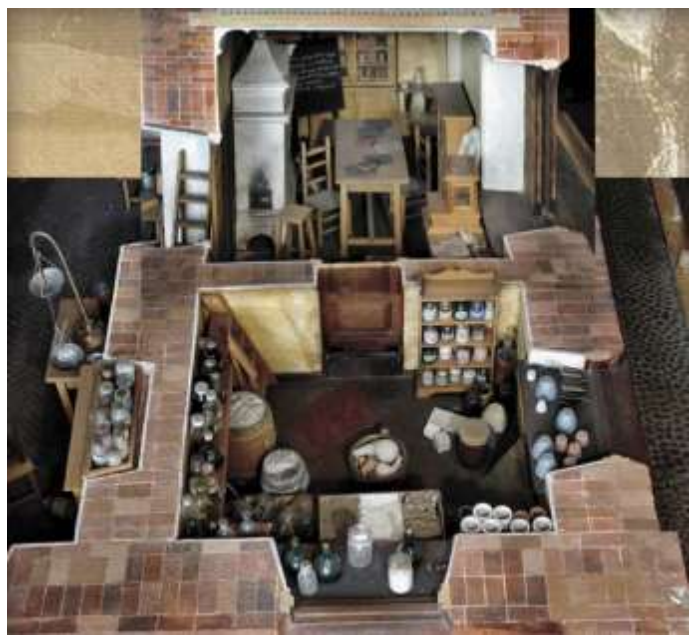

Fig.3: Reconstructed 1:10 model of the Lomonosov's chemical lab contains 480 miniature replicas of instruments and tool according to the 1759 list of the lab equipment. (a - above) In the center of the hall, on the brick foundation aredistillation and melting furnaces, iron assay furnace, round glass-making and square enamel furnaces. (b - bottom) detail: on the left - laboratory scales, background - room for writings and lectures, forefront – chemicals storage room  (R.I.Kaplan-Ingel, 1948; Museum of M.V.Lomonosov, St.Petersburg; photo courtesy M.Khartanovich [15]).

The construction budget of over 1000 rubles (some 0.5M$ today) does not seem astronomical by today's standards, but it was substantial when the Academy's total state support was about 25,000 rubles and the total budget (including income from the print house and various workshops) was often not sufficient to support the salaries of over 100 staff members. Lomonosov, for example, had an annual salary of 360 rubles as an adjunct and 600 rubles as a Professor since fall'1745, but was often paid quarterly by books printed by the Academy (it was his job to sell them to meet his salary).

On October 12, 1748, Lomonosov reported to the Academy of Sciences that "The laboratory, which was founded on the 3rd of August at the Botanical Garden, has been brought to completion with all the external and internal structure"[16]. Just two months of construction culminated seven years of struggle from the "project need statement" to "start of research operation." The delay was due to several significant societal and natural events.

**Kunstkamera Fire**

One of the biggest financial challenges of the first years of the Razumovsky's presidency – which had a direct impact on the laboratory construction plans – was the devastating fire of 1747. On December 5, 1747, at about five o'clock in the morning, a fire broke out in the building housing the Kunstkamera. The source of the fire was found to be located in the tower at the west wing of the gallery, where various offices and "physical cabinet" rooms were located. The wooden tower in the middle of the building, which housed the observatory with its instruments and large artful "Gottorf globe" (a kind of planetarium), was also burned out completely. Lomonosov, who saw the fire, believed that a malfunction of the Academy's stoves or chimneys was the cause of the disaster. He and Richmann later documented damage to the physics room and instruments destroyed by the fire. Fearing that the fire would spread to the entire building, the employees threw books, documents, and other things out of the windows into the snow, and a significant part of the Kunstkamer exhibits during the night commotion was plundered. A special decree was issued afterward requiring the return of things to the Academy of Sciences. But the strain on the Academy budget was enormous, and another reason why Lomonosov had to look to the Senate rather than the Academy for funding for the lab.

**Schumacher Affair**

Another cause for delay was the "Schumacher affair," sparked by growing clashes between patrons of the lab.

Academics at the time, as well as other state servants, generally had to depend on powerful protectors to be promoted. These patrons could either be individuals or factions aligned with factions at the Imperial Court[17]. Factions at the Academy were "French", "German" and "Russian" parties (though the factions were fluid, for there were Russians in the German faction, etc), centered around J.N.Delisle, J.D.Schumacher and Andrey Nartov, respectively. The dynamics and infighting of these factions were determined by the court winds, which were chaotic after the enthronement of Elizaveta Petrovna on December 6, 1741, with growing animosity between "Russians" and "French" on one side and "Germans" on the other that crystallized around J. D. Schumacher.

Early in 1742, Andrey Nartov (1683—1756, a military engineer and inventor, a personal craftsman of Peter I of Russia, and from 1735 a member of the Academy of Science as a head of the Academy's lathe workshop) filed two reports to Senate about fraud and mismanagement in the Academy, accusing their long-time enemy J.D.Schumacher as the principal villain. Shortly before, Schumacher had illegally arrested a foreign student, Tobias Kenigsfeldt - Delisle's closest collaborator, who had just returned with him from an expedition to Siberia - for refusing to sign on to Schumacher's accusation to the Senate that Delisle and Leonhard Euler had deliberately delayed work on the Atlas of Russia. The Senate ordered Schumacher to release the arrested man and let him go back to Germany – but, enraged, Schumacher filed another denunciation to the Senate on July 2, 1742 against all three: Delisle, Euler and Kenigsfeldt.

In response to the accusations filed by Delisle, Nartov and a number of the Academy staff, Empress Elizaveta Petrovna signed a decree on November 30, 1742 appointing an investigative commission over Schumacher, who was himself arrested, and Nartov was entrusted with the leadership of the Academy. Nartov appointed Lomonosov, still an adjunct, to take control over Academy archives, library and Kunstkamer. The Senate commission, headed by Prince Bosris Yusupov (1695-1759) did not find any violations, and by the end of December 1742, Schumacher was released. At the beginning of 1743, the Senate ordered Schumacher compensated for the time spent during his arrest. Schumacher's friend Jacob Staehlin, member of the Academy and with connections at the Imperial Court, surely played a role in these events. The commission completed the investigation a year later, finding Schumacher guilty only of using Academy's wine for his own use, for which he was fined 109 rubles and 38 kopecks.

The accusers did not get off so easy. While Delisle and Nartov were returned to their positions relatively quickly, Lomonosov got in serious trouble. He had been too zealous in his control of the Academy's archives, and routinely and rudely denied access to them, even to the Academy's secretary Winsheim (*Christian Nicolaus von Winsheim*, 1694—1751). Lomonosov was particularly rude on several occasions. Once, while drunk (according to Winsheim), Lomonosov scolded the professor because he was not native Russian, and was not fluent enough in Latin; using a surly Russian threat, Lomonosov threatened "to straighten up" Winsheim's teeth. Several Academicians complained about Lomonosov's behavior to the Yusupov's commission, which then summoned Lomonosov for an interrogation – but he refused, claiming that he answered to no one but Andrei Nartov. The commission moved swiftly and arrested Lomonosov on June 4, 1743. Lomonosov was eventually put under house arrest for almost 7 months (one of the most scientifically productive periods of his life, he later noted), and his salary was halved, a punishment that was a particular hardship as it came shortly before the arrival in St. Petersburg of his wife Elizabeta-Christine from Germany, their daughter, and her brother). Lomonosov was ultimately released and exonerated after a public apology delivered to the Academic Conference on January 27, 1744 in the form prescribed by the Senate Commission[18]. The other signers of Nartov's complaint (commissioner M.S.Kammer, translators I.S.Gorlitsky and N.I.Popov, cleck D.Grekov, copiers V.Nosov and I.Pukhort, gravure pupil A.Polyakov, students P.Shishkarev, C.Starkov and M.Kovrin) were sentenced to punishment with whips and *batogs*, but were pardoned and allowed to return to the Academy.

The war between Schumacher and the academicians did not stop until Count Kirill Grigoryevich Razumovsky (1728-1803) assumed the title of President of the Academy in May 1746. New complaints to the Senate followed, but again yielded nothing and the academicians could do nothing to undermine Schumacher's authority. Razumovsky began to put things in order at the academy. The intelligent and insinuating adjunct Grigory Teplov, who had great influence on Razumovsky, soon became an assessor of the academic office, began to take part in its management. However, Schumacher managed to cultivate Teplov, a fellow student of Lomonosov, and essentially remained in control of the Academy, and via a series of clever administrative moves managed to force J.N.Dlisle to leave the Academy in June 1747).

For Lomonosov, a positive fallout of these clashes was his election to full Professorship (Academician) in August 1745[19]. This was partly a compromise between various parties at the Academy. The promotion gave him greater clout both in internal Academic affairs and finally allowed him traction in his efforts to initiate construction of the chemical lab.

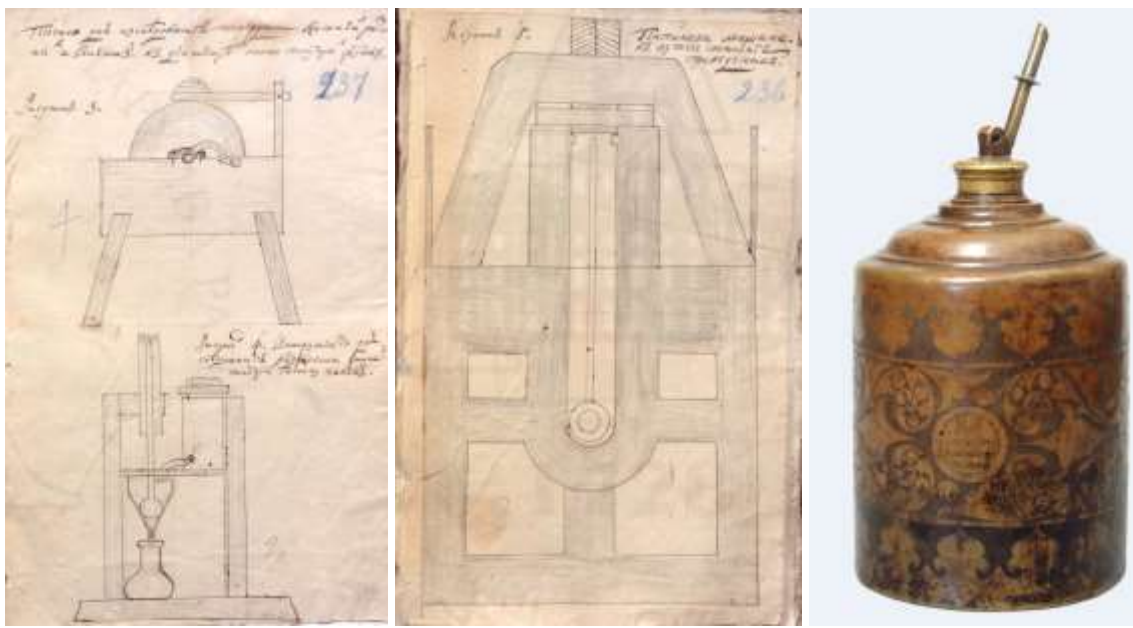

Fig.4: Lomonosov's lab instruments for chemical and physical studies: (left to right) Lomonosov's own hand drawings of a grinder machine for hardness determination, a tool for viscosity measurements by counting number of drops, "improved Papin machine" (high pressure device, May 1752, from collection of the St.Petersburg Branch of the Archive of Russian Academy of Sciences), and his distillation cube made of copper (master F.Kiselev, Demidov's Ural workshops, 1748; State History Museum, Moscow).

**Equipment and Staff**

Lomonosov began purchasing and collecting equipment for his lab soon after the decision to proceed. About the end of July, 1748, he submitted to the Chancellery a list of "items needed for the chemical laboratory", tools, chemical vessels and materials, and Lomonosov also indicates where all those can be obtained[20]. Most of the items were delivered from the factories by February 1749, though some of the ordered instruments were not manufactured even by 1752. Eventually, the lab housed almost five hundred items and instruments (according to the 1757 description; Fig. 3b). In its center were nine types of furnaces and ovens used for melting, distillation, glass-making, assaying, roasting, heating with a bath, and others. The furnaces were placed on a low platform, between four pillars supporting the arch. A free passage was left around so that one could easily observe the fire. There were dozens of large and small retorts, flasks, recipients, bottles of white and green glass, evaporating cups, funnels, mortars, cans with various chemicals and reagents ranging from the simplest to the most complex.

A report submitted to the Academy in May 1752 gives an idea of the variety of types of instruments and tools used by Lomonosov for research in chemistry and physics – a dozen thermometers, as many as nine weights and scales including precise one with half a mg accuracy), a microscope, a tool for studying the viscosity of liquid materials, tools for studying the rigidity of solids by pressure and breaking, high pressure vessels, a vacuum pump ("antliya"), pyrometer, apparatus for determining specific gravity, and autoclaves [21]. Figure 4 shows some of them: a) a grindstone with a diameter of 1.5 feet to study the hardness of minerals and glasses clamped in a side wooden holder, and an instrument for studying the viscosity of liquid materials by the number of drops. This consisted of a funnel filled with a liquid coming from a reservoir (upper right); the position of a glass ball on long stem inserted into the funnel regulates speed of the liquid outflow of a liquid, so that the number of drops falling from a funnel in a certain period of time indicates of the viscosity of the liquid; b) an "improved Papin's machine" used to obtain relatively high pressures (ordered by Lomonosov and manufactured at the Sestroretsk factories, invented the French physicist Denis Papin (1647-1713) ; c) a "distillation cube" - a 4-liter capacity cylindrical copper vessel with a screwed copper cover and a copper tube soldered at an angle. This was ornamented with large leaves and curled stems and inscribed "M V Lomonosov Academia St. Piter-Burch 1748".

Effective operation required an experienced laboratory staff, butLomonosov had a long and uneven success with laboratory workers. The first was Petr Pryanishnikov, who was hired in the Fall 1748 and fired in February 1749 due to laziness and criminal deeds. He was followed by Johann Maneke who was hired in April 1749 but left for Mecklenburg in June 1751. Franz Bettinger took over in June 1751 and lasted until May 1756, when Lomonosov's student Vasily Klementiev took over until his untimely death in February 1759.

Bettinger's 1752-1756 logbooks[22] give an impressive account of what it took to operate the lab. A budget of some 80 to 100 rubles a year was used to purchase up to 100 bags of coal and 20 *sazhen's* of wood (about 120 m$^3$) for furnaces and ovens, buy some 90 different chemicals – from simple things such as fractured glass, ice and alabaster to Mg, Hg, Pb, Zn, As, niter, strong and light acids, etc. – and about a hundred pieces of labware and vessels made of glass, ceramics, metals and wood. In total, 94 suppliers – ranging from "blacksmith Petr Milyukov, 30 kopecks" to "goldmeister Joahan Palm, 5 rubles 50 kopecks" – and 24 contractors were involved in the laboratory operation.

Besides the laboratory workers, whose annual salaries ran up to 200 rubles, the lab was staffed with two stokers "to carry and load wood and coal into ovens, wash and clean dishes, prepare, rub and grind material and keep the entire laboratory clean"[23].

**Applied studies, scientific research, education, outreach.**

Lomonosov worked in the laboratory for almost 8 years until 1757, and by his own account he carried out there more than four thousands tests and experiments. The laboratory journal for 1751 (the only one that has survived) contains about 20 series of experiments, some of these including up to 100 experiments[24]. While most of Lomonosov's biographers and scholars emphasize his pioneering development of mosaics, he conducted a broad spectrum of studies and trained many students, an activity that contributed to the lab's subsequent impact on Russian science. N. Raskin's authoritative study of the lab provides detailed accounts of numerous analyses of ores, minerals and salts carried out following per requests sent to the Academy from various sources, including: studies of dyes of both natural and lab origin (such as developing a method for manufacturing *Prussian Blue* more cheaply than the imported product); improving the technology of silicates and porcelain studying the composition and melting of optical glasses including flint-glass for telescopes; studies of hydrates of metal oxides which were often tied to the searches of vivid glass colors;, exploratorations of recipes and technological processes for obtaining alloys for metal mirrors of telescopes; and pyrotechnical works as needed for making fireworks[25].

*Glasses, smalts and mosaics*

Lomonosov's studies of silicates and their reactions, particularly the preparation of colored glasses, started almost as soon as he moved into the laboratory, and soon became a preoccupation that lasted till the end of his life [26]. He achieved significant scientific and technological success in that area, and effectively used it to promote his position at the Imperial Court hierarchy and satisfy his entrepreneurial zest. Mid-20th century glass technologist Nikolay Kachalov considered Lomonosov's lab – with 6 out of nine furnaces dedicated to silicates and full of related dedicated equipment - as a cradle of the "modern science of glasses, covering broad spectrum of most complex of topics", and particularly emphasized its experimental methods: adherence to truly scientific techniques of experimental research, including strict control of experimental conditions, accurate description of observed phenomena, systematic storage of samples, and maintenance of a laboratory journal[27].

Lomonosov's general procedure was to glass melts from materials, which were carefully weighed out on a sensitive balance with 1 grain (0.0625 g) accuracy, and then to add mineral pigments. Those were usually prepared by precipitation, by dissolving a metallic salt in water and adding ammonia or potassium hydroxide, which would precipitate a hydrous oxide or basic salt of copper, iron, mercury, or other heavy metal. These precipitates were dried and added to the molten glass together with other organic dyes or potassium nitrate, for color modification. Depending on the pigments used, base components (usually domestic sands) and conditions of the fusion, Lomonosov was able to obtain a wide variety of colored glasses and (nontransparent) smalts. These included "an excellent green, of grassy color, very like an emerald", "liver color," "beryl red," "very similar to excellent turquoise but semitransparent," "light purple," and so on – all in all, over 100 of colors and over a thousand of shades. Of particular note and acclaim was Lomonosov's rediscovery of a ruby glass, the secret of which had been lost, with use of gold compounds. These experiments resulted in a wider range of colored glasses than were available to the mosaic artists of western Europe. Lomonosov's works were known abroad, and in a letter of March 30, 1754, his colleague Leonhardt Euler wrote from Berlin, "I congratulate you on having produced glass of all possible colors. Our chemists consider this a great discovery."[28]

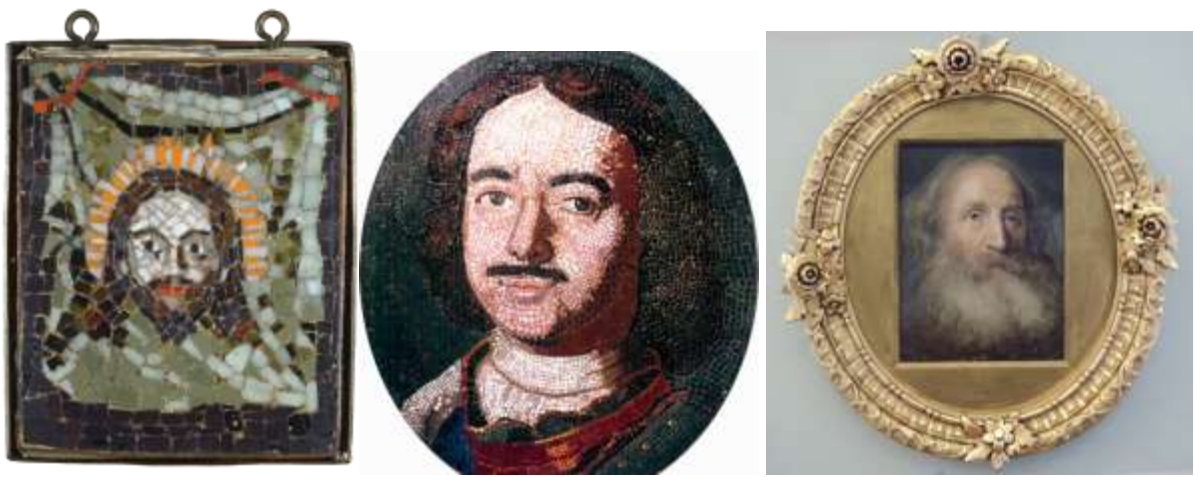

Fig.5: Lomonosov's early mosaics made at the chemical lab: (a- left) Savior "Not-Made-by-Hand", Icon of the Mandylion of Edessa; engraving on the back of the copper pan: "*This Mandylion image of Christ our Savior, at the request of the most illustrious Countess Mavra Yegoryevna Shuvalova, was composed by Mikhail Lomonosov at the beginning of experiments in mosaic art in St. Petersburg in 1753*." (smalts, mosaics, 11x9 cm; State History Museum, Moscow) ; (b – center) portrait of Peter the Great, 1753 (38X33 cm, State Russian Museum, Saint Petersburg); (c – right) for comparison – one of the most latest mosaic portrait of an old man, 1768 (Factory of M.V. Lomonosov, craftsman Matvey Vasiliev, 1768; 40x31 cm, State Russian Museum, Saint Petersburg).

In 1750 Lomonosov turned to mosaics, reportedly following an encouragement from his patron, Vice Chancellor Count Mikhail I. Vorontsov (1714-1767). Vorontsov had in his St. Petersburg home two glass mosaic paintings brought from the Vatican workshops: a portrait of Empress Elisabeth I made by Alessandro Cocchi, the official mosaicist of Pope Benoit XI, who sent it to the Empress as a present; and "Weeping Apostle Peter" by Guido Reni. Having already developed by that time an impressive palette of 112 different colors, reportedly even larger than the palette of the Vatican, Lomonosov was set to revive the art of mosaics, which was largely forgotten in Russia after despite magnificent achievements in the 11th and 12th centuries, such as the ancient mosaic decorations in the Saint Sophia Cathedral in Kyiv and at the Michael the Archangel Cathedral in Novgorod.

Lomonosov's mosaics greatly improved his reputation in the Imperial Court. Of Lomonosov's 40 mosaics, (23 of which are preserved), the first three were made in the chemical laboratory in 1752-1753[29]. The first, a Madonna (after one by the Italian artist Solemene, which no longer exists) was about 300 square and of quality and was given to the Empress Elizaveta Petrovna, who rewarded the ambitious scientist and placed the mosaic among the icons in her apartments.

As the Office of the Academy of Sciences noted on September 24, 1752:

On the 15th of this month, the Councilor and Professor Lomonosov, in a report to the Office, communicated that he had the highest mercy on September 4 to present to Her Imperial Majesty the mosaic image of the Mother of God he had compiled. The said image was compiled from the original of the glorious Roman artist Solimen, and made of more than 4000 of all the composite pieces, all made by his hands, and 2184 experiments were made in a glass furnace to make the compositions"[30].

The other two mosaics made at the lab [Figs. 5a and 5b] were of rougher composition. The 11 x 9 cm image of the Savior "Not-Made-by-Hands" is rather course. It was assembled in 1753 at the request of Countess Mavra Shuvalova (1708-1759), the closest personal friend of the Empress and sister-in-law of Lomonosov's major patron Ivan Shuvalov. In a letter to her husband, Count Petr Shuvalov, on May 10, 1753, Lomonosov asked him to forgive the imperfections of the mosaic, expressed his intention to hone his skills, and mentioned that he had requested the Senate to allow him to establish a mosaic factory: "After establishment of the factory," he wrote, "I have no doubt that in a short time I will achieve perfection in this art". Chancellor Vorontsov and the Shuvalovs pulled strings, and on December 16, 1752, by decree of the Senate, Lomonosov was allowed to "start a bead factory", which was assigned to the department of the Manufacture Collegium. At the same time, he was given a loan of 4,000 rubles, got a monopoly privilege on production of the assorted colored-glass items for 30 years, and on March 15, 1753 obtained ownership over a sizable estate - the entire village of Ust-Ruditsa, some 60 km from St. Petersburg - with 211 "souls" (male serfs with families) needed for the factory operation.

After that, the art of the mosaics in Russia greatly improved [Fig. 5c], thanks largely to Matvey Vasil'ev (1731-1782) and Efim Mel'nikov (1734-1767), students whom Lomonosov had trained in the chemical laboratory. We will cover Lomonosov's mosaics enterprise in greater detail in subsequent articles.

*Physico-chemical studies, conservation of matter in chemical reactions*

The prominent Russian chemist Nikolai Menshutkin (1874-1938) considered Lomonosov to have founded "physical chemistry" over 120 years before Wilhelm Ostwald, to whom its founding is routinely attributed.[31]. "Physical chemistry," Lomonosov wrote, "is a science that explains, on the basis of the provisions and experiments of physics, what happens in mixed bodies during chemical operations"[32]). In his explanations, Lomonosov largely rejected the then-dominant theories of "imponderable substances" such as phlogiston and applied mostly mechanical, corpuscular views and theories. For example, his "New theory of colors" – the result of analysis of his own studies in the lab – assumed that the three main colors red, yellow and blue corresponded to the oscillatory motions of corpuscles of three different sizes and shapes,– the larger ones making red and smaller ones blue[33]. We will consider Lomonosov's scientific breakthroughs and intuitions on variety of topics in subsequent articles, and consider here only one episode directly related to his experiments on calcination (oxidation) of metals carried out in the lab, and commonly considered as an important step (if not direct experimental confirmation) toward the law of conservation of matter in chemical reactions.

As well known, the English chemist Robert Boyle broke with earlier explanations of fire to propose a corpuscular account. In 1673 he conducted an experiment as follows: a piece of lead was placed in a hermetically sealed glass retort, and weighted. Then it was heated in a fire until the lead turned to cinder, the retort opened (whereupon Boyle noted the whistling sound of inrushing air) and the lead weighed again. He considered the lead's increase in weight as a proof that the substance of fire was capable penetrating through the glass of the retort and uniting with the metal.  In 1756, Lomonosov repeated that experiment himself, and found out the that when air is not admitted into the vessel (i.e., the report remained sealed after heating and cooling), the total weight of the vessel and its contents remained constant – indirectly suggesting the general law of conservation as it applies to the total weight of chemically reacting substances. This experiment was reported to the Academy of Sciences[34], but the results were never published. The Soviet scholar of Lomonosov, Yakov G. Dorfman proposed that Lomonosov had not published the results because he realized an alternate explanation: that his poor vacuum pump (estimated to keep the vacuum pressure only at 15-20 mm Hg) might have leaked in enough air to make the total weight of the vessel before and after firing constant. While this may be correct, we can only add two reservations: a) while experiments on calcination of metals in vacuum were in the original Lomonosov's plans, no results of such are known; b) even with imperfect pumping, the results of the pumping-sealing-weighting-heating-weighting experiment would depend on the duration of the heating as the oxidation reaction would be slowed by the deficiency of oxygen.

In 1773, the French chemist Antoine Lavoisier, repeated Boyle (and Lomonosov's) experiment and converted tin into "lime" (oxide) in a hermetically sealed vessel with air in by heating the metal with a large magnifying glass. The total weight of the vessel with tin, after the transformation of tin into "lime", remained unchanged. Lavoisier also found that the amount of air taken after the experiment is reduced by 1/5 and that the remaining air does not support combustion and respiration, and described his results in "Opuscules physiques et chimiques" (1773). That was the end of the phlogiston concept. Henry Leicester has investigated whether Lomonosov's results and theoretical works and corpuscular ideas on the subject were known to Lavoisier and concluded that while it is "probably correct… that Lavoisier knew of the work of Lomonosov, which was widely commented upon at the time it was published, he probably would not have recognized its importance for his own thinking… Therefore it is not surprising that he was not mentioned by the French scientist".[35]

*Education and outreach*

According to the then-latest Charter of the Academy (1747), its mission included education of researchers, which was the duty of its Academic University. The Academy had salaried staff positions of "Academics" (members of the Academic conference, with 9 academicians, 9 adjuncts and 10 honorary members) and "Professors" (6 in total, who, supervised up to 30 students). Most Professors were Academicians who delivered lectures. For example, in 1749, the core science lectures were given by academicians Georg Richmann in physics, Lomonosov in chemistry, Joseph Adam Braun (1712-1768) in philosophy, etc. After taking these basic courses the students were supposed to specialize. A student's standard salary was 100 rubles per year compared with an adjunct's 360 rubles, a professor's 600 rubles and 860-1200 rubles for academicians), which was not particularly high given how expensive life was in St. Petersburg. That, plus the absence of prospects for Academy students to get into the civil service rank system, with its guaranteed promotions and pensions, made attracting talented students into the Academy a challenge.

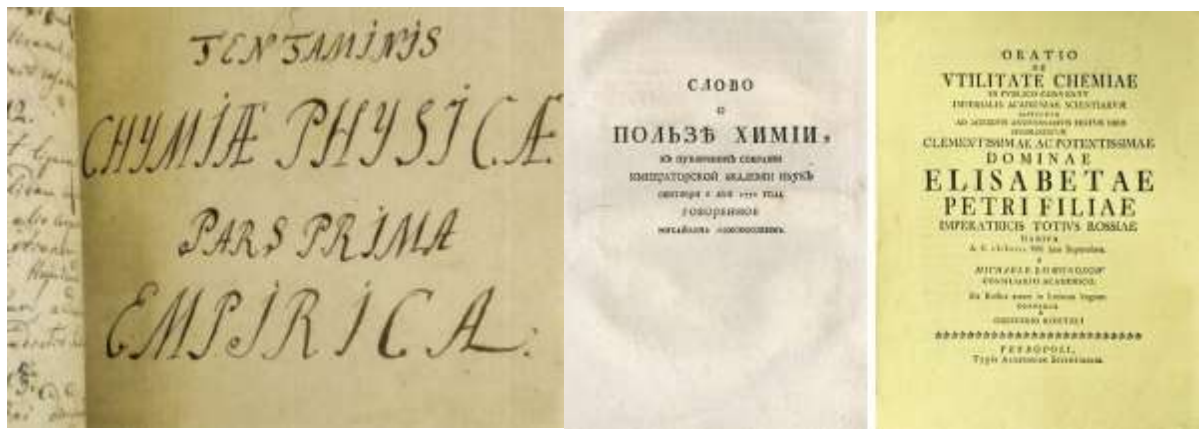

Fig.6: (a, b,c - left to right) Lomonosov's; title pages of the "Word on the benefits of chemistry in the public collection of the Imperial Academy of Sciences, September 6, 1751 spoken by Mikhail Lomonosov", published in Russian (1751) and Latin (1759) (Russian National Library, St. Petersburg; source: Source: N. A. Kopanev ''Knizhnii Oformitel' Mikhailo Lomonosov: k priumnozheniyu pol'zi i slavi Otechestva'' [''Book Illustrator Mikhailo Lomonosov''], *Nauka iz Pervikh Ruk [Science First Hand]* **4** (40) (2011), 54–65, website https://scfh.ru/files/iblock/35f/35feb4fffd1af0c6835469c21d3a9807.pdf)

The education and training of students was a very important part of the newly created chemical laboratory. On January 15, 1749, Lomonosov reported to the Academy: "I, the undersigned, announce that in the first third of this year 1749 <...> I will do chemical experiments in the Chemical Laboratory to study minerals and other things and will show the first foundations of chemistry to the students"[36]. But the first students were "assigned" to chemistry classes only in February 1750, when, later that year, Lomonosov started "to teach practical chemistry at the Academic Laboratory in the afternoon hours on Mondays and Thursdays". Besides presenting true experimental chemistry methods, partly following what he had learned from *Johann* Henckel and other chemists encountered in Germany, Lomonosov intended from the very beginning to focus on physical foundations and methods in chemistry:

> In my chemical lectures, which I should read to student youth, I consider it very useful to add, where possible, physical experiments to chemical experiments…. Therefore, in the entire course of experimental chemistry compiled by my labors, it is necessary 1) to determine the specific gravity of chemical bodies; 2) to investigate the adhesion between their particles: a) by breaking bodies, b) by squeezing , c) grinding on a bar, d) counting drops of liquid; 3) describe the figures of crystalline bodies; 4) expose bodies to the action of Papin's machine; 5) observe degrees of heat everywhere; 6) examine bodies, especially metals, by grinding. In a word, to test everything that can only be measured, weighed and determined by calculation.[37]

Lomonosov's lectures on physical chemistry ended in May 1753 and, as far as is known were his last activity as a university professor.

All existing records indicate at least six students attended his courses: M. Sofronov, I.N. Fedorovsky, N.Popovsky, I. Bratkovsky, S. Rumovsky, and V. Klementiev. Among those only the last two entered a scientific career – future Academician Stepan Rumovsky (1734-1812)[38] in physics, math and astronomy, and Vasily Klementiev (1731-1759) in chemistry (both of whom were singled out by Lomonosov for their progress). Born in Moscow in the family of a deacon, Klementiev studied at the Slavic-Greek-Latin Academy and in 1748 – following footsteps of Lomonosov himself 12 years earlier - was transferred to the Academic University. Due to his poor knowledge of the Latin, he was first assigned to the Academic Gymnasium, then as a University student he attended Lomonosov lectures untill 1753. Under Lomonosov's his guidance he wrote his dissertation "On the increase in weight received by some metals after deposition", approved by the Academic Assembly in 1754[39]. In 1756, at the request of Lomonosov, Vasily was appointed to the vacant post of laboratorian of the Chemical Laboratory where he worked till his untimely death in 1759 at the age of 27[40]. Klementiev also left a handwritten synopsis of Lomonosov's lectures "Introduction in True Physical Chemistry" [Fig. 6a].

Since its establishment, the Academy of Science arranged lectures for the public, usually around Imperial family jubilees or other important court events. These lectures - open to all and known for their attendance by courtiers – were carefully planned to keep the public broadly informed about the latest developments and news in science. Lomonosov was one of the most active lecturers. One of his first outreach public speeches - "Oration on the Usefulness of Chemistry" [Figs.6b and c] delivered on September 6, 1751 on the "long-awaited name day of Her Imperial Majesty the most powerful and invincible Grand Empress Elizaveta Petrovna Autocrat of All Russia" - was directly related to his laboratory studies and to general views on physical foundations of chemistry[41].

**Aftermath and impact**

Lomonosov worked in his laboratory until 1757, though less frequently after the foundation of the Ust' Ruditsa mosaic factory in 1753. By that time he had gained prominent and financial well-being, and – together with his wife, daughter Elena (born February 21, 1749 in St. Petersburg) and brother-in-law Johannes Zilch – he moved to a large estate built for him on the bank of Moika River (at the present address of Bolshaya Mosrkaya, 61). There, among other arrangements and labs, he set up a small home laboratory for auxiliary experimental research[42].

After Lomonosov, the chemical laboratory was headed by several Professors of Chemistry: Ulrich Christopher Salchow (1722-1787, lab director in 1757-1760), Johann-Gottlob Lehman (1700-1767), but who did very little for the laboratory and died in 1767, and Erich Laxman (1737-1796). The latter lived the Academy owned apartments in "Bonov's house" where Lomonosov had previously lived, next to the laboratory which he took over in 1770 but chemistry was just one of his many interests among zoology, economy, meteorology, botany and zoology. In October 1776, the laboratory was raided by robbers, and in September 1777 it was badly damaged by a flood, so Laxman had to devote a lot of time to its restoration. The laboratory was restored in the 1780s, when it was taken over by an adjunct and then academician Nikita Sokolov (1748-1795) who revived Lomonosov's traditions, started a course of public lectures on chemistry in the laboratory, illustrating them with numerous experiments. In 1792, at the request of Sokolov, the director of the Academy of Sciences Ekaterina Dashkova (1743-1810)[43] released him as head of the chemical laboratory. This position was entrusted to the then adjunct of chemistry Yakov Zakharov (1765-1836).

In 1793, the site, which included the chemical laboratory, the residential building in which Lomonosov once lived, and a plot of land of the 2nd Liniya of Vasilievsky ostrov, belonging to the Academy of Sciences, was sold to Academician Nikolay Ozeretskovsky (1750-1827). He rebuilt the former laboratory, renovating it into housing but leaving old walls of the laboratory and its foundations intact. Subsequently, the building of the laboratory and the wooden Bonov's House in which Lomonosov had lived were owned by several different people. During the terrible times of the Siege of Leningrad in the Great Patriotic War of 1941-1945, the already many-times-rebuilt former chemical laboratory and the nearby dilapidated wooden house were destroyed. The USSR and later the Russian Academy of Sciences made several attempts to reconstruct Lomonosov's first chemical lab,[44] in 1985 and in the early 2000's, but without success. The site and relatively well-preserved underground remnants of the lab foundation are now the property of the Russian State Pedagogical University named after A. Herzen, and local enthusiasts are attempting to reconstruct the lab for the tercentennial of the Russian Academy of Science in 2024-25.

Menshutkin has emphasized that Lomonosov's chemical laboratory was the first ever of the kind to combine research and teaching, followed only in 19th century by Professor Liebig in Giessen in 1825[45]. Though without an international profile, the laboratory was locally significant. Not only that it was the birthplace of the Lomonosov's many ideas, which eventually elevated him to the status of a "Russian National Myth"[46], but it also became a symbol that Russian scientists appealed to as a model, and it led to the establishment of other chemical laboratories being established, such as the similar one built at Moscow University in 1755[47].

Lomonosov, as it happens, was also the first great Russian poet of modern times, and inevitably he wrote poetry about his achievements at the laboratory. Several lines of the following poem, "Letter on Usefulness of Glass,"[48] addressed to his patron and friend Ivan Shuvalov, are frequently cited in Russian textbooks:

> *Those thinking wrong about things, Shuvalov,*
> 
> *Which glass revere below minerals,*
> 
> *Whose shining rays allure the eyes -*
> 
> *As beautiful and useful is the glass;*
> 
> *And sing I praise before you in delight*
> 
> *Not to expensive stones, not to gold, but glass.*"

Aside from his skills at mosaics and fireworks, Lomonosov also used his poetic gifts to promote the cause of his laboratory!

**Acknowledgments**

Authors wish to express sincere gratitude to the group of enthusiasts actively working on the project of restoration of the Lomonosov's first chemical laboratory in St. Petersburg. Dr. Victor Borisov and Prof. Sergei Bogdanov of the Russian State Pedagogical University named after A.Herzen, Lead Scientist of NPO "Science for Construction" Victor Korenzvit and Acad. Sergei Lyulin of the Russian Academy of Sciences hosted one of us (V.S.) during the 2021 visit to St.Petersburg, toured to the place of the lab and provided many useful details on the laboratory operation and equipment and their plans to re-build it for the purpose of scientific and historical outreach.

---

[1] For Lomonosov's early years to 1730, see R. Crease and V. Shiltsev, "Pomor Polymath: The Upbringing of Mikhail Vasilyevich Lomonosov, 1711–1730," *Physics in Perspective* **15** no. 4 (2013): 391–414.

[2] For Lomonosov's education in Germany, see R. Crease and V. Shiltsev, Fueling Peter's Mill: Mikhail Lomonosov's Educational Training in Russia and Germany, 1731–1741", *Physics in Perspective* **20** no.3 (2018): 272-304.

[3] K.Hufbauer, *The formation of the German chemical community, 1720-1795* (Univ of California Press; 1982)

[4] P.M. Luk'yanov, "The First Chemical Laboratories in Russia," *Chymia* (Annual Studies in the History of Chemistry, H.M.Leicester, ed.; Philadelphia :Univ. Pennsylvania Print) **9** (1964): 59-69 at p. 65.

[5] M.Raskin, *Khimicheskaya laboratoriya M.V. Lomonosova* (Moscow, Leningrad: AN SSSR Publishers, 1962) (in Russian: Раскин Н.М. Химическая лаборатория М.В. Ломоносова. М.; Л., 1962.) pp. 136–138.

[7] H.G.Good, "On the early history of Liebig's laboratory", *Journal of Chemical Education,* **13** no.12 (1936): 557.

[8] P. Pekarskiy, *Istoria Imperatorskoi Akademii Nauk in Petersburg, t.II [History of Imperial Academy of Sciences in Petersburg, vol.II]* (St.Petersburg: Imper. Acad. Sci. Print, 1872) 343

[34] Lomonosov, *Complete Works* (ref. 8), 10: 392

[35] H.M.Leicester, "Boyle, Lomonosov, Lavoisier, and the Corpuscular Theory of Matter." *Isis* **58**.2 (1967): 240-244.

[36] Lomonosov, *Complete Works* (ref. 8), 10: 377

[37] Report in Latin to the Conference of the Academy of Sciences, filed on May 11, 1752, in Lomonosov, *Complete Works* (ref. 8), 9: 55-57.

[38] G.E.Pavlova, *Stepan Yakovlevich Rumovsky, 1734-1812* (Mooscow: USSR Academy of Sciences, Nauka, 1979); also Y.Balashov, "Rumovsky, Stepan Yakovlevich" in T.Hockey, et al., eds. *Biographical Encyclopedia of Astronomers* (New York: Springer, 2014): 1876-1877.

[39] Biography of V.Kelmentiev and full text of his dissertation (originally in Latin, Russian translation) can be found in N.M.Raskin *Vasily Ivanovich Klementyev, student and laboratory assistant of M.V. Lomonosov* (Moscow, Leningrad: USSR Academy of Sciences, 1952).

[40] Klementiev died February 23, 2759 at his home, after an epileptic accident, but the origin of that is not clear – Raskin (ref.36 above) connects it with poisoning by chemicals, while B.Menshutkin cites drinking (B,N,Menshutkin, *Works of Lomonosov on Physics and Chemist*ry(Moscow, Leningrad : USSR Academy of Sciences, 1936): 352 and 448).

[41] Lomonosov, *Complete Works* (ref. 8), 2: 365-369; English translation in H.M.Leicester, tr. ed., *Mikhail Vasileivich Lomonosov on corpuscular theory,* (Cambridge, MA: Harvard Univ. Press, 1970): 186-202.

[42] V.K.Makarov, "Lomonosov's home chemical laboratory" in *Lomonosov: Collection of articles and materials*, v.III (Moscow, Leningrad: Academy of Sciences, 1951): 347–349.

[43] See on her, e.g., R.P.Crease, "The pioneer princess." *Physics World* 31.3 (2018): 22.

[44] V.A.Korenzvit, "On Archelogical studies at the location of Bonov's House and Lomonosov's chemical laboratory" in *Lomonosov: Collection of articles and materials*, v.X (St.Petersburg: Nauka, 2011): 417.

[45] B.N.Menshutkin, *Russia's Lomonosov* (Princeton, NJ: Princeton University Press, 1952): 130

[46] See e.g. S.Usitalo, *The Invention of Mikhail Lomonosov: A Ruyssian National Myth* (Boston: Academic Studies Press, 2013).

[47] See in P.M. Luk'yanov, "The First Chemical Laboratories in Russia" (ref.4): 68

[48] That very poetic and lengthy "Letter on the Usefulness of Glass, to the Most Excellent Mr. General Lieutenant, the Her Imperial Majesty Kammerger, the Curator of Moscow University

and the Orders of the White Eagle, St. Alexander and St. Anna Cavalier Ivan Ivanovich Shuvalov" was written in 1752, see Lomonosov, *Complete Works* (ref. 8), 8: 508-522;